\documentstyle[emulateapj]{article}

\lefthead{Livio, Ogilvie, \& Pringle}
\righthead{Extracting energy from black holes: the relative importance of
the Blandford-Znajek mechanism}

\begin{document}

\title{Extracting energy from black holes: the relative importance of the
Blandford-Znajek mechanism}
\author{M.~Livio$^{1}$, G.~I.\ Ogilvie$^{1,2}$ and J.~E.\ Pringle$^{1,2}$}
\affil{$^1$Space Telescope Science Institute, 3700 San Martin Drive,
 Baltimore, MD 21218, USA}
\affil{$^2$Institute of Astronomy, Madingley Road, Cambridge CB3 0HA, UK}
\affil{E-mail: mlivio@stsci.edu; gogilvie@ast.cam.ac.uk; jep@ast.cam.ac.uk}

\slugcomment{Accepted for publication in the Astrophysical Journal}

\begin{abstract}

We critically assess the role of the Blandford-Znajek mechanism in the
powering of outflows from accretion disk-fed black holes. We argue
that there is no reason to suppose that the magnetic field threading
the central spinning black hole differs significantly in strength from
that threading the central regions of the disk. In this case, we show
that the electromagnetic output from the inner disk regions is
expected in general to dominate over that from the hole. Thus the spin
(or not) of the hole is probably irrelevant to the expected
electromagnetic power output from the system. We also point out that
the strength of the poloidal field in the center of a standard
accretion disk has been generally overestimated, and discuss scenarios
which might give rise to more significant central poloidal fields.

\end{abstract}

\keywords{accretion, accretion disks -- black hole physics --
galaxies: nuclei, active, jets -- magnetic fields}

\section{Introduction}

Blandford \& Znajek (1977) demonstrated an interesting process by
which the spin energy of a black hole might be extracted by magnetic
fields supported by a surrounding accretion disk.  Over the last
twenty years this process has assumed prime importance as being widely
believed to be the major mechanism that powers radio jets in active
galactic nuclei (e.g.\ Begelman, Blandford, \& Rees 1984; Rees,
Begelman, Blandford, \& Phinney 1982; Blandford 1991, 1993; Rawlings
\& Saunders 1991, Wilson \& Colbert 1995; Moderski \& Sikora 1996a), and
 elsewhere (Paczy\'nski 1998).  Recently, however, Ghosh \& Abramowicz
(1997) have called into question the concept that the Blandford-Znajek
process can provide the primary power output from an accretion
disk-fed black hole, and have argued that the strength of the magnetic
field threading the black-hole horizon has been widely overestimated.
In this paper, we take the critical assessment of Ghosh \& Abramowicz
(1997) a step further.  We put forward the case that (a) even if the
Blandford-Znajek mechanism is operating, its power output is, in
general, dominated by the electromagnetic power output of the inner
regions of the disk, and (b) for a standard, thin accretion disk the
dominant power output is that due to viscous heating in the disk
itself.

\section{Electromagnetic/wind and disk power outputs}

We here consider the mechanisms by which a poloidal magnetic field
threading the disk or the hole can manage to extract energy from the
disk or hole.  This energy can be extracted in the form of Poynting
flux (i.e. purely electromagnetic energy) or in the form of a
magnetically driven material wind.  With this in mind, we keep the
discussion fairly general.

\subsection{Output from the hole: the Blandford-Znajek mechanism}

Blandford \& Znajek (1977) showed that if a spinning black hole is
threaded by an externally generated magnetic field, then spin energy
can be extracted from the hole.  The mechanism can be understood in
terms of a simple electromagnetic analogy (e.g.\ Thorne, Price, \&
MacDonald 1986), according to which the black hole acts as if it has a
magnetic diffusivity, $\eta_{\rm h}$, which is such that the decay
time-scale for a field on the scale $R_{\rm h}$ of the hole, $t_{\rm
decay}\sim R_{\rm h}^2/\eta_{\rm h}$, is approximately equal to the
light crossing time $\sim R_{\rm h}/c$.  This implies that the black
hole acts as if it has an effective diffusivity given by
\begin{equation}
\eta_{\rm h}\sim R_{\rm h}c~~.
\end{equation}
Then, if the hole is threaded by a poloidal field of magnitude $B_{\rm
ph}$, and the hole is spinning with angular velocity $\Omega_{\rm h}$,
the maximum toroidal field that can be generated at the horizon is
\begin{equation}
B_{\phi{\rm h}}{\rm(max)}\sim\left({{R_{\rm h}\Omega_{\rm
h}}\over{c}}\right)B_{\rm ph}~~.
\end{equation}
The rate at which work is done by the hole on the external medium is
given by
\begin{equation}
L_{\rm BZ}\sim\left({{B_{\rm ph}B_{\phi{\rm
h}}}\over{4\pi}}\right)(R_{\rm h}\Omega_{\rm h})\pi R_{\rm h}^2~~.
\end{equation}
Thus the maximal rate at which energy can be generated, which is
usually taken to be the Blandford-Znajek luminosity, is the rate
corresponding to when $B_{\phi{\rm h}}$ is maximal, and is given by
\begin{equation}
L_{\rm BZ}{\rm(max)}\sim\left({{B_{\rm ph}^2}\over{4\pi}}\right)\pi
R_{\rm h}^2\left({{R_{\rm h}\Omega_{\rm
h}}\over{c}}\right)^2c~~.\label{lbzmax}
\end{equation}
Whether or not such a maximal rate can be achieved in reality depends
crucially on how the field lines are loaded with material, that is in
effect on how much inertia they have.  This is often discussed in
terms of `impedance matching' between the resistance of the hole and
the material surrounding the hole (e.g.\ MacDonald \& Thorne 1982;
Thorne, Price, \& MacDonald 1986).

The above analysis was carried out under the assumption (Blandford \&
Znajek 1977) that the far ends of the field lines threading the hole
are tethered somewhere at large distance in a medium with zero angular
velocity.  If, instead, the field lines were tethered in the disk,
with angular velocity $\Omega_{\rm d}$, then the above formula for
$L_{\rm BZ}{\rm(max)}$ would still be valid provided that $\Omega_{\rm
h}^2$ is replaced by $4 \Omega_{\rm d}(\Omega_{\rm h}-\Omega_{\rm d})$
(MacDonald \& Thorne 1982; Ghosh \& Abramowicz 1997).  In this case
the energy cannot be going into a jet, but is, if $\Omega_{\rm
h}>\Omega_{\rm d}$, being used to spin up the material in the disk at
the other end of the field line.  Conversely, if $\Omega_{\rm
h}<\Omega_{\rm d}$, the hole is being spun up by the disk (Moderski \&
Sikora 1996b).  It may be that in reality the hole is threaded by a
mixture of open field lines, which give rise to a wind or jet, and
closed field lines, which are attached to the inner disk and which
transfer rotational kinetic energy of the hole to rotational kinetic
energy of the disk in a process exactly analogous to that occurring in
the magnetosphere of a rapid rotator (Ghosh \& Lamb 1978; Livio \&
Pringle 1992).

\subsection{Electromagnetic output from the disk}

By similar considerations to those presented in the previous section
it is evident that the electromagnetic/wind output from the disk can
be written in the form
\begin{equation}
L_{\rm d}\sim\left({{B_{\rm pd}B_{\phi{\rm d}}}\over{4\pi}}\right)(\pi
R_{\rm d}^2)R_{\rm d}\Omega_{\rm d}~~.
\end{equation}
Here $B_{\rm pd}$ and $B_{\phi{\rm d}}$ are the poloidal and toroidal
components, respectively, of the field at the disk surface, and
$\Omega_{\rm d}$ is the angular velocity in the disk at a relevant
radius, $R_{\rm d}$. Since we are interested in the radii where most
of the disk energy is available we shall take $R_{\rm d}$ to be a
radius close to the inner disk edge, within which, say, half of the
accretion luminosity is emitted. Thus we may take $R_{\rm d}$ to be a
factor of a few times larger than $R_{\rm h}$.

Because the disk has a much higher conductivity than the hole, the
maximal value of $B_{\phi{\rm d}}$ is of the same order of magnitude
as $B_{\rm pd}$ (cf. Livio \& Pringle 1992).  In fact $B_{\phi{\rm
d}}$ would reach a much higher value if it were limited only by the
magnetic diffusivity (e.g.\ Campbell 1987); the limit here arises from
the instability (or reconnection) of a predominantly toroidal field
(e.g.\ Biskamp 1993).  As above, whether or not such a value of
$B_{\phi{\rm d}}$ can be achieved depends on how mass is loaded on to
the poloidal field lines.  Thus the maximal disk wind luminosity is
\begin{equation}
L_{\rm d}{\rm(max)}\sim\left({{B_{\rm pd}^2}\over{4\pi}}\right)\pi
R_{\rm d}^2\left({{R_{\rm d}\Omega_{\rm
d}}\over{c}}\right)c~~.\label{ldmax}
\end{equation}

\subsection{Comparison of electromagnetic disk and hole outputs}

From the above analysis we may now compare the relative outputs from
the disk and the hole. From equations (4) and (6), we see that
\begin{equation}
{{L_{\rm BZ}{\rm(max)}}\over{L_{\rm d}{\rm(max)}}}\sim\left({{B_{\rm
ph}}\over{B_{\rm pd}}}\right)^2\left({{R_{\rm h}}\over{R_{\rm
d}}}\right)^2\left({{c}\over{R_{\rm d}\Omega_{\rm d}}}\right)a^2~~,
\end{equation}
where $a\sim R_{\rm h}\Omega_{\rm h}/c$ is the dimensionless spin
parameter of the hole ($0<a<1$).  Writing approximately that $(R_{\rm
d}\Omega_{\rm d}/c)\sim(R_{\rm h}/R_{\rm d})^{1/2}$ , we find that
\begin{equation}
{{L_{\rm BZ}{\rm(max)}}\over{L_{\rm d}{\rm(max)}}}\sim\left({{B_{\rm
ph}}\over{B_{\rm pd}}}\right)^2\left({{R_{\rm h}}\over{R_{\rm
d}}}\right)^{3/2}a^2~~.
\end{equation}
Since $a^2<1$ and $R_{\rm d}$ is a factor of a few times $R_{\rm h}$,
it is evident that unless $B_{\rm ph}$ is significantly larger than
$B_{\rm pd}$ the electromagnetic/wind luminosity extracted from the
hole by the Blandford-Znajek mechanism is less than, or at best
comparable to, the electromagnetic/wind luminosity extracted by
similar electromagnetic considerations from the disk.

\section{The poloidal magnetic field}

In the original paper by Blandford \& Znajek (1977), it was
acknowledged that if the poloidal fields threading the inner disk and
the hole were comparable then the Poynting flux from even a maximally
spinning black hole would be only a fraction of the comparable flux
from the disk.  Since then, however, the concept of a {\it
`fiducial'\/} field accumulated steadily on the hole by a net influx
of magnetic field dragged inwards by the accretion flow has been
introduced (e.g.\ Begelman, Blandford, \& Rees 1984; Thorne, Price, \&
MacDonald 1986; Blandford 1993).  The `fiducial' field is estimated by
assuming that the hole is accreting at the Eddington limit, that
matter is falling freely radially inwards, and that the resulting
kinetic energy density of the material is equal to the local magnetic
energy density.  This `fiducial' field should be regarded as a severe
upper limit to the field threading the hole, since a field of such
magnitude would be able to terminate, or significantly restrict, the
radial accretion.  Nevertheless, on the basis of the fiducial field,
large luminosities have been claimed for the Blandford-Znajek
mechanism to the extent that it is now widely regarded as a {\it sine
qua non\/} for jet formation in active galactic nuclei. The history of
the various estimates of the poloidal field threading the hole has
been recently documented by Ghosh \& Abramowicz (1997), and we do not
intend to repeat all the arguments here.  We shall however need to
touch upon some of the salient points.

In order to build up a poloidal field on the hole that substantially
exceeds the poloidal field threading the inner disk two physical
processes must be able to occur. First, the disk must be able to
transport poloidal field radially inwards. And second, such a field
through the hole must be maintainable at the inner edge of the disk.
The second process has been addressed at length by Ghosh \& Abramowicz
(1997).  They argue simply that, because any poloidal field threading
the hole acts back to repel any poloidal field being inwardly advected
by the disk, the strength of the field threading the hole cannot be
much higher than the strength of the field threading the inner
disk. This argument is strengthened by the realization that the
currents that generate the field threading the hole must be situated
in the disk rather than in the hole, and thus that the field through
the hole is just a continuation of the field through the disk (see
also Ghosh \& Abramowicz 1997).

In the remainder of this section, we discuss whether these two
processes can occur in a standard, thin accretion disk.

\subsection{Advection of poloidal field: standard accretion disk}

There is a considerable literature on the launching of
electromagnetically driven jets/winds from the surface of an accretion
disk, under the assumption that the disk is threaded by a suitably
configured poloidal field (e.g.\ Blandford \& Payne 1982; K\"onigl
1989; Pelletier \& Pudritz 1992), but little discussion of how the
field configuration was set up in the first place.  The hope appears
to be that a general external poloidal field will be advected inwards
by the accretion flow in the disk.  However, it has been demonstrated
that in a standard viscously driven accretion disk, unless the ratio
of magnetic diffusivity to viscosity (the inverse magnetic Prandtl
number) in the disk is very small (comparable to $H/R$), such inward
advection of field does not take place (Lubow, Papaloizou, \& Pringle
1994a; Reyes-Ruiz \& Stepinski 1996; Heyvaerts, Priest, \& Bardou
1996).  It is important to note, however, that in a disk in which the
angular momentum transport is mainly due to the effects of
self-sustaining hydromagnetic turbulence (Balbus \& Hawley 1991;
Hawley, Gammie, \& Balbus 1995, 1996; Stone, Hawley, Gammie \& Balbus
1996; Brandenburg, Nordlund, Stein, \& Torkelsson 1995), the magnetic
Prandtl number is likely to be of order unity (Parker 1971; Pouquet,
Frisch \& L\'eorat 1976; Zel'dovich, Ruzmaikin, \& Sokoloff 1983;
Canuto \& Battaglia 1988).  Thus, if the disk surrounding the black
hole is at any radius a standard accretion disk in which the dominant
mode of angular momentum loss is by viscous transport within the disk,
poloidal magnetic flux cannot be simply advected inwards from
infinity.

\subsection{Strength of the poloidal field: standard accretion disk}

Under the assumption that the inflow of material to the black hole is
from a standard Shakura \& Sunyaev (1973) accretion disk, we now
consider the likely strength of the poloidal field in the inner
regions.  Ghosh \& Abramowicz (1997), while noting that the usual
assumption is that the field strength is given by equating the
magnetic pressure $P_{\rm mag}$ to the maximum pressure in the disk,
$P_{\rm d}$ (e.g.\ Moderski \& Sikora 1996b), draw attention to the
fact that the relevant field is the one produced by dynamo processes
in the disk, and that current numerical simulations indicate that the
limiting value of this field $B_{\rm dynamo}$ is such the the magnetic
pressure $P_{\rm mag}$ is only a small fraction of the gas pressure in
the disk.  Using this estimate for the field strength, coupled with
the argument that the field threading the hole cannot greatly exceed
this, enables them to conclude that the power of the Blandford-Znajek
process has been seriously overestimated.  However, it should also be
noted that a magnetic dynamo mechanism in an accretion disk will tend
to produce magnetic field primarily on scales of the order of the disk
thickness, $H$, whereas in the computations in the Ghosh \& Abramowicz
paper they make the assumption (inspired by the fact that in the boxes
used in magnetic dynamo simulations $H/R\sim1$; see their Figure~1)
that the fields so produced have typical length-scales of order
$R_{\rm d}$.  Indeed, in the solution they present for computing the
field threading the hole, they make the assumption that the field
threading the disk has a configuration such that $B_R/B_z=7/3$.
However, if, as they assume, this is indeed an accretion disk with
magnetic Prandtl number of order unity, then $B_R/B_z$ should not
exceed a value of order $H/R$ (Lubow et~al.  1994a).  We emphasize
that it is important to make the distinction between the mean
(large-scale) field and the {\it rms\/} (small-scale) fields.

In reality, it seems likely that a dynamo mechanism in an accretion
disk might be able to give rise to more global fields with typical
length-scales larger than $H$, but the processes by which it can do so
are as yet largely unexplored in numerical simulations (see, however,
Armitage 1998). The reason for this is that the simulations have so
far been limited to a small element of the disk, and are incapable of
considering the disk as a global entity.  Tout \& Pringle (1996) have
suggested that outside the main body of the disk, the internal dynamo
might be able to produce magnetic loops mainly on scales of order $H$,
which could then interact by differential rotation and reconnection to
produce an inverse cascade to larger length-scales (see also Romanova
et~al. 1998).  For the particular idealized case which they
considered, they found that $B(\lambda)\propto\lambda^{-1}$, where
$B(\lambda)$ is the flux density at scales of $\lambda$ or greater.  A
similar scaling might be expected from any dynamo process operating
through an inverse cascade. In this case, we would expect the size of
the large-scale field threading the disk, $B_{\rm pd}$, to be given
approximately by
\begin{equation}
B_{\rm pd}\sim\left({{H}\over{R_{\rm d}}}\right)B_{\rm dynamo}~~. 
\end{equation}
Noting that the energy dissipated in the disk by the dynamo mechanism
is given approximately by
\begin{equation}
L_{\rm acc}\sim\left({{B_{\rm dynamo}^2}\over{4\pi}}\right)(2\pi
R_{\rm d}\cdot 2H)(R_{\rm d}\Omega_{\rm d})~~,
\end{equation}
where the first term in parentheses represents the magnetic stress in
the disk, we see that the ratio of the maximal electromagnetic flux
from the disk (eq.~[6]) to the energy generated in the disk is
approximately
\begin{equation}
{{L_{\rm d}{\rm(max)}}\over{L_{\rm
acc}}}\sim \left({{R_{\rm d}}\over{H}}\right)\left({{B_{\rm
pd}}\over{B_{\rm
dynamo}}}\right)^2~~,
\end{equation}
which, using the above estimate, gives
\begin{equation}
{{L_{\rm d}{\rm(max)}}\over{L_{\rm acc}}}\sim{{H}\over{R_{\rm d}}}~~.
\end{equation}
This estimate is consistent with the observations of jets and winds
produced from a variety of objects (e.g.\ Pringle 1993; Livio
1997). Equation (9) therefore indicates that in the case of a standard
disk even the calculation of Ghosh \& Abramowicz (1997) overestimates
the power of the Blandford-Znajek process.

\section{Discussion}

We have seen from the above that in order for the Blandford-Znajek
mechanism to dominate the power output it is necessary for the
magnetic flux on open field lines threading the black hole to greatly
exceed the magnetic flux on open field lines threading the inner
regions of the disk. We have also seen that, if the flow surrounding
the black hole is that of a standard accretion disk, this does not
come about.  We now consider two cases of accretion flow, which are
not standard thin accretion disks, in which inward advection of
poloidal field is more likely to occur.

\subsection{Advection-dominated accretion flows}

In an advection-dominated accretion flow (ADAF), the basic idea is
that energy released in the accretion process is not radiated locally
but is, rather, retained by the fluid as internal energy and advected
into the hole (see, for example, Narayan \& Yi 1995; and the review by
Svensson 1998).  As far as the present discussion is concerned, the
major difference between this kind of accretion flow and the standard
disk is that the disk is geometrically thick in the sense that $H\sim
R$.  The accretion is driven by viscous processes with $\alpha\sim1$,
and hence with $v_R\sim v_\phi$.  What this implies (see Section 3.2)
is that the inward flow velocity is comparable in magnitude to the
outward diffusion velocity for a poloidal field threading the disk.
This means that there could in principle be some non-negligible radial
advection of poloidal flux.  While it might be possible to set up a
steady configuration in which inward advection of poloidal flux is
balanced at each radius by outward diffusion, there is no reason to
expect that the field threading the hole (which is in any case
generated by currents in the disk) can significantly exceed the field
threading the inner disk.

\subsection{Non-standard angular momentum loss}

It is evident that if we wish to produce significant advection of
poloidal flux to the inner disk regions it is necessary to ensure that
the radial inflow velocity in the disk exceeds the radial diffusive
outflow rate of poloidal field.  Since this cannot be done using a
standard disk in which the inflow is due to outward diffusion of
angular momentum through the disk, it follows that we need to look for
other mechanisms for outward transport of angular momentum.

\subsubsection{Gravitational torques}

It the disk is self-gravitating, as is thought to occur in the early
stages of protostellar disks, and in the outer regions of disks around
galactic nuclei, then non-axisymmetric instabilities can give rise to
significant outward transport of angular momentum (Paczy\'nski 1978;
Boss 1984; Anthony \& Carlberg 1988; Lin \& Pringle 1987, 1990;
Sellwood \& Lin 1989; Laughlin, Korchagin, \& Adams 1997).  Since such
a process is not driven by hydromagnetic instabilities it is
conceivable that the magnetic Prandtl number might be quite different
from unity, and that significant inward transport of poloidal field
might be able to take place.  Although the inner regions of disks
around black holes either in AGN or in Galactic binaries are not
usually considered to be self-gravitating, there might be an
interesting exception here if one considers the disk generated in the
dynamical disruption of a neutron star by a black hole, which occurs
in some models for $\gamma$-ray bursts (Rasio 1996; M\'esz\'aros \&
Rees 1997; Paczy\'nski 1998).

\subsubsection{Inflow driven by magnetic winds}

It has been argued by a number of authors (e.g.\ Blandford \& Payne
1982; Pudritz \& Norman 1986; K\"onigl 1989; Pelletier \& Pudritz
1992; Lovelace, Romanova, \& Contopolous 1993) that a magnetically
driven disk wind might be the main mechanism by which excess angular
momentum is removed from disk material, and so might be the main
mechanism that drives an accretion disk.  Here again the inflow
velocity can in principle significantly exceed any outward diffusion
rate for poloidal field, especially if the poloidal field is strong
enough to suppress the Balbus-Hawley instability.  If such a mechanism
were able to give rise to a steady state, it would be necessary to
appeal to some process (such as the interchange instability; Spruit \&
Taam 1990; Lubow \& Spruit 1995; Spruit, Stehle, \& Papaloizou 1995)
that counterbalances the steady inward dragging of poloidal field and
allows outward diffusion of field to occur.  Thus, as in Section
4.2.1, it is envisaged that a steady poloidal field configuration is
set up, with inward advection and outward diffusion producing a
balance and a steady gradient in the poloidal field.  However, for the
reasons discussed above, there is no reason to expect such physical
processes to give rise to a poloidal field threading the hole that is
significantly enhanced over the poloidal field threading the inner
disk.

In addition, the idea that such a steady balance can be set up at all
has been brought into question (Lovelace, Romanova, \& Newman 1994;
Lubow, Papaloizou, \& Pringle 1994b; Agapitou \& Papaloizou 1998).
The main point here is that the process of wind removal of angular
momentum occurs locally and directly at each radius in the disk.  An
annulus in the disk which succeeds in getting rid of angular momentum
to a wind does not require the presence of neighbouring annuli to do
so.  Thus different annuli which manage to dispose of their angular
momentum in this way are to some extent independent dynamical
entities.  Furthermore, efficient removal of angular momentum from a
particular annulus leads to inward movement, outward bending of
poloidal field lines, and consequently enhanced wind outflow, enhanced
removal of angular momentum, and further inflow (Lubow, Papaloizou, \&
Pringle 1994b).  (This instability may be tempered by the fact that
strong bending of field lines impedes an outflow by making the disk
sub-Keplerian; Ogilvie \& Livio 1998.)  However, even if such unstable
wind driven accretion occurs (and at least in the disks in cataclysmic
variables there is evidence that it does not, Livio 1997), there is no
particular reason to suppose that at any stage the strength of the
poloidal field threading the hole is significantly greater that the
strength of the poloidal field threading the inner disk, except
possibly for brief dynamical interludes.  Thus, here again, it seems
difficult to set up a credible picture in which electromagnetic
extraction of spin energy from the hole dominates in a steady, or even
a time-averaged sense, over electromagnetic extraction of spin energy
from the disk material.

\section{Conclusions}

Blandford \& Znajek (1977) noted that if the poloidal magnetic field
threading the black hole is comparable in strength to the poloidal
field threading the inner parts of the accretion disk, then the black
hole's contribution to the electromagnetic output (that is, the output
due to the Blandford-Znajek effect) is likely to be ignorable. We have
argued here that it is hard to conceive of a situation in which the
magnetic field threading the hole is significantly stronger than the
field threading the inner disk. This comes about for two main
reasons. First, the currents which generate the field must, of
necessity, be in the disk, and not in the hole; and second, the hole
is (in effect) a very poor conductor, compared to the surrounding disk
material. We conclude, therefore, that independent of the spin of the
black hole, the electromagnetic output of the disk (in the form of
Poynting flux or magnetically driven wind) dominates that from the
hole.

In addition, we have argued that the poloidal field strengths in the
centers of standard accretion disks (e.g.\ Moderski \& Sikora 1996b;
Ghosh \& Abramowicz 1997) have been overestimated in the
literature. We have pointed, however, to some accretion scenarios in
which the poloidal field strength could be significantly enhanced with
respect to the standard disk picture.

\section*{Acknowledgments}

This work was started at the programme on `The Dynamics of
Astrophysical Discs' held at the Isaac Newton Institute for
Mathematical Sciences, Cambridge. We thank Ralph Pudritz, Pranab Ghosh
and Martin Rees for useful discussions. GIO and JEP thank the Space
Telescope Science Institute for hospitality. ML acknowledges support
from NASA Grant NAG5-6857.

\end{document}